\documentclass[twocolumn,american,english]{revtex4-1}
\usepackage[T1]{fontenc}
\usepackage[latin9]{inputenc}
\usepackage{geometry}
\usepackage{color}
\usepackage{babel}
\usepackage{float}
\usepackage{amsmath}
\usepackage{amssymb}
\usepackage{graphicx}
\usepackage[unicode=true,pdfusetitle,
 bookmarks=true,bookmarksnumbered=false,bookmarksopen=false,
 breaklinks=false,pdfborder={0 0 1},backref=false,colorlinks=true]
 {hyperref}
\hypersetup{
 linkcolor=green}

\makeatletter
\PassOptionsToPackage{hyphens}{url}\usepackage{hyperref}

\makeatother

\begin{document}

\title{Study of the likelihood of Alfvénic mode bifurcation in NSTX and
predictions for ITER baseline scenarios}

\author{V. N. Duarte}
\email{vduarte@pppl.gov}

\address{Princeton Plasma Physics Laboratory, Princeton University, Princeton,
NJ, 08543, USA}

\address{Institute of Physics, University of São Paulo, São Paulo, SP, 05508-090,
Brazil}

\author{N. N. Gorelenkov}

\address{Princeton Plasma Physics Laboratory, Princeton University, Princeton,
NJ, 08543, USA}

\author{M. Schneller}

\address{Princeton Plasma Physics Laboratory, Princeton University, Princeton,
NJ, 08543, USA}

\author{E. D. Fredrickson}

\address{Princeton Plasma Physics Laboratory, Princeton University, Princeton,
NJ, 08543, USA}

\author{M. Podestà}

\address{Princeton Plasma Physics Laboratory, Princeton University, Princeton,
NJ, 08543, USA}

\author{H. L. Berk}

\address{Institute for Fusion Studies, University of Texas, Austin, TX, 78712,
USA}

\date{\today}
\selectlanguage{american}%
\begin{abstract}
Rare Alfvénic wave transitions between fixed-frequency and chirping
phases are identified in NSTX, where Alfvénic waves are normally observed
to exhibit either chirping or avalanching responses. For those transitions,
we apply a criterion {[}Duarte et al, Nucl. Fusion 57, 054001 (2017){]}
to predict the nature of fast ion redistribution in tokamaks to be
in the convective or diffusive nonlinear regimes. For NSTX discharges
in which the transition is not accompanied by changes in the beam
deposited power or modifications in the injected radiofrequency power,
it has been found that the anomalous fast ion transport is a likely
mediator of the bifurcation between the fixed-frequency mode behavior
and rapid chirping. For a quantitative assessment, global gyrokinetic
simulations of the effects of electrostatic ion temperature gradient
turbulence and trapped electron mode turbulence on chirping were pursued
using the GTS code. The investigation is extended by means of predictive
studies of the probable spectral behavior of Alfvénic eigenmodes for
baseline ITER cases consisting of elmy, advanced and hybrid scenarios.
It has been observed that most modes are found to be borderline between
the steady and the chirping phases.
\end{abstract}
\maketitle
\selectlanguage{english}%

\section*{Introduction}

\selectlanguage{american}%
Energetic-particle-driven Alfvénic instabilities can seriously degrade
the performance of present-day and next-generation fusion devices
\cite{HeidbrinkSadler1994,LauberReview2013,Gorelenkov2014,ChenZoncaRevModPhys.2016,Heidbrink2008}.
The control of these instabilities is, therefore, considered essential
for the ITER performance \cite{JacquinotITERExpertGroup1999,Gorelenkov2014}.
To avoid and mitigate instabilities due to fast ions, it is important
to understand what is the nature of the induced transport (e.g., coherent
prompt losses \cite{ChenXiPRL2013PrompLosses}, convective losses
due to phase-space structures \cite{Heidbrink1995Chirping} and diffusive
losses due to phase-space stochastization \cite{GarciaMunozPRL2010})
and the associated spectral character of the instability \cite{DuarteAxivPRL}.
For this purpose, in this paper we investigate what conditions delineate
the transition of Alfvénic modes between fixed-frequency and chirping
phases in NSTX. In addition to that, typical ITER cases are analyzed
and predictions are made regarding the likelihood of modes to chirp
or to oscillate steadily at a nearly constant frequency. 

From the theory perspective, the onset of chirping has been linked
with the relative importance between stochastic and coherent processes
affecting the resonant population \cite{BerkPRL1996,BerkPLA1997,Lilley2009PRL,Lilley2010}.\foreignlanguage{english}{
Experimentally, several elements have been identified as altering
the Alfvén wave spectral behavior, such as radiofrequency (RF) waves
\cite{MaslovskyMauelPRL2003,MaslovskyPoP2003,Fredrickson2015NF},
background plasma beta \cite{GryaznevichSharapovPPCF2004}, beam beta
\cite{Heidbrink1995Chirping}, 3D fields \cite{BortolonPRL2013} and
rotational transform \cite{MelnikovNF2016vol56numb11,MelnikovNF2016vol56numb7}.
}Recently, the rare emergence of chirping in DIII-D has been shown
to be related to a marked decrease of the inferred fast ion micro-turbulence
levels \cite{DuarteAxivPRL}. 

In Refs. \cite{DuarteAxivPRL,DuartePoP2017}, theoretical predictions
for realistically computed tokamak modes were compared with experiments.
It was observed, both in the theory and in the experiment, that fast
ion micro-turbulence is a mediator between mode nature transition
for several typical tokamak scenarios, while exhibiting little macroscopic
anomalous transport \cite{Pace2013}. The fact that ion anomalous
diffusity is typically much smaller in spherical tokamaks (STs) than
in conventional tokamaks has been proposed \cite{DuarteAxivPRL} as
the explanation for the longstanding observation that chirping is
more common in STs relative to conventional tokamaks. A criterion
proposed to distinguish between the two typical scenarios (chirping
vs fixed-frequency), which was shown to be sensitive to the relative
strength of scattering (from collisions and micro-turbulence) and
drag processes, ultimately translates into a condition for the applicability
of reduced quasilinear modeling for realistic tokamak eigenmodes.
Recently, DIII-D had dedicated experiments \cite{VanZeeland2017IAEA}
to stress-test the chirping prediction \cite{DuarteAxivPRL}. The
experiments employed negative plasma triangularity as a means to decrease
turbulence levels. In those shots, chirping was much more prevalent
than in the usual and more turbulent oval or positive triangularity
cases. In addition, chirping was also observed for modes located around
internal transport barriers, even in positive triangularity. From
the numerical side, turbulence stochasticity has been recently included
in a bump-on-tail simulation \cite{WoodsDuarteNF2018}, which showed
its effect on chirping suppression.

The purpose of this paper is twofold. First, we investigate whether
the conclusion of chirping based on turbulence holds for NSTX, where
ion micro-turbulence is already observed to be low (as compared to
ion neoclassical transport \cite{KayeNF2007}), or whether there are
other elements that determine the mode nonlinear evolution. This is
addressed by means of global gyrokinetic simulations. Chirping is
a major issue in connection with fast ion losses in tokamaks and there
is currently no understanding of how to systematically avoid them
in NSTX-U. The second goal of this work is an attempt to anticipate
what will be the probable spectral nature of toroidicity-induced and
reversed shear Alfvén eigenmodes (TAEs and RSAEs) in ITER. The profiles
employed in this study are fiducially constructed using a transport
code. 

This article is organized as follows. In Sec. II, we present gyrokinetic
analysis of Alfvénic mode bifurcation in NSTX and its comparison with
theoretical predictions. Sec. III is devoted to a study of the possibility
of occurrence of mode chirping in ITER baseline scenarios (reversed
shear, hybrid and elmy H mode) and Sec. IV presents discussions and
conclusions.

\section*{Analyses of rare transitions between constant frequency and chirping
in NSTX}

\selectlanguage{english}%
A distinctive feature of Alfvénic wave behavior in STs \cite{McClementsFredrickson2017}
as compared to conventional tokamaks is that chirping and avalanches
are customary in the former and infrequent in the latter. The ubiquitous
Alfvénic chirping in NSTX is observed to be a precursor of the phase
locking of high-intensity modes of several toroidal mode numbers,
known as avalanche \cite{Podest2012,Fredrickson2009PoP}. During the
avalanche, the escape of a substantial fraction of fast ions occurs,
which can typically reach up to $40\%$. Wave chirping has been identified
to transition to avalanches in NSTX only when the fast particle energy
was greater than about 30\% of the total plasma energy \cite{FredricksonNF2014},
with higher chances of avalanches to appear correlating with lower
values of \foreignlanguage{american}{maximum energetic ion speed divided
by the Alfvén speed} \cite{McClementsFredrickson2017}. The conditions
determining the likelihood of the chirping-to-avalanche transition
are therefore, to a certain extent, experimentally identified. We
therefore turn our attention to the likelihood of transition from
a steady frequency to chirping instead.

Typically in NSTX, Afvénic modes are already chirping when they first
appear in a spectrogram, although mode intensity can vary considerably.
The constant frequency phase is normally not observed. Eventually
chirping modes undergo conversion to the avalanche phase as more beam
power is injected. It is often challenging to identify cases in which
they appear in their fixed-frequency phase.\foreignlanguage{american}{
From an extensive NSTX database, we have selected rare steady-to-chirping
transitions, with no influence of RF or 3D fields,} in order to test
the proposed interpretation that fast ion micro-turbulence can be
a determining factor behind the transition.\foreignlanguage{american}{
Although rare, these transitions offer a unique testbed to decipher
the parameters that need to be varied to induce change in the nonlinear
character of modes. For some cases, we have observed that chirping
and steady phases appear to co-exist for different Alfvénic modes,
which possibly indicate that each mode has its own threshold to undergo
transition between the two phases. From the initial set of transitions,
we observe that in most cases the modes switch over to chirping because
of an increase in the applied beam power, with the spectral change
happening within the characteristic fast ion slowing down time. This
increases the strength of the drive and likely allows the modes to
access a harder nonlinear phase. This observation is in agreement
with Ref. \cite{Heidbrink1995Chirping}, that reported a correlation
between chirping and high fast ion pressure. The chances of chirping
increasing at higher drive may be related to a convective amplification
mechanism \cite{ZoncaConvectiveAmplificationNF2005} and also be related
to mode structure deformation due to the presence of EPs, both of
which are beyond the scope of this work. Therefore, to obtain a comparable
set of steady-to-chirping transitions, we further analyze only a sub-set
of cases with roughly constant beam beta throughout the transition.}
Only three discharges satisfied this constraint, among which only
two allowed for a successful equilibrium reconstruction: NSTX shots
$\#128453$ and $\#135388$.

\selectlanguage{american}%
\begin{widetext}

\begin{figure}[H]
\selectlanguage{english}%
\begin{centering}
\includegraphics[scale=0.55]{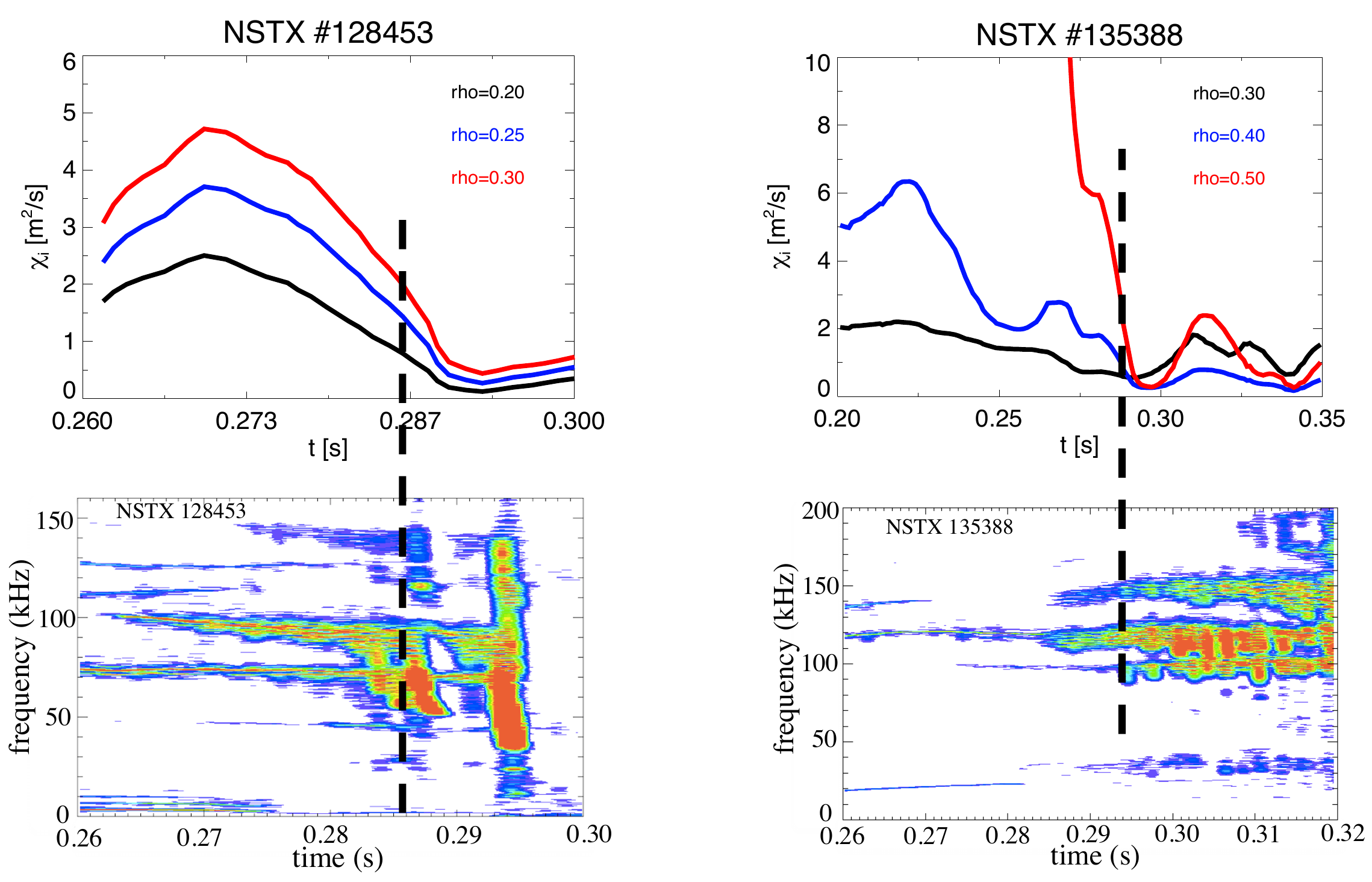}
\par\end{centering}
\selectlanguage{american}%
\caption{\foreignlanguage{english}{Correlation between the onset of Alfvénic chirping (lower plots) and
an improvement of thermal ion confinement, as inferred from TRANSP
(upper plots), for NSTX pulses $\#128453$ and $\#135388$.\label{fig:TRANSPCorr}}}
\end{figure}

\end{widetext}

\selectlanguage{english}%
Previous chirping analysis \cite{DuarteAxivPRL,DuartePoP2017} relied
on transport coefficients calculated by TRANSP for thermal ions. The
fast ion coefficients were then scaled from the thermal ones using
the relations reported in Ref. \cite{ZhangLinChen2008PRL,Hauff2009PRL,PueschelNF2012}.
These are based on an analytic approach and numerical examination
of a simulation database. One can appreciate in Fig. \ref{fig:TRANSPCorr}
the correlation between the emergence of chirping and the decrease
of thermal ion turbulent levels, which can be used as a proxy for
the fast ion anomalous diffusivity \cite{Heidbrink2009PRL}. The values
of $\chi_{i}$ in Fig. \ref{fig:TRANSPCorr} are unusually high before
the start of the chirps. In the present work, however, we analyze
new cases more quantitatively consistent by performing individual
gyrokinetic simulations at given time slices of the mode evolution,
taking into consideration the experimental plasma parameters for each
individual discharge. This allows us to understand what type of turbulence
is dominant and offers insights on its key characteristics, such as
the origin of its drive.

The global turbulence simulations reported in this study are carried
out using the \textsc{Gyrokinetic Tokamak Simulation} (\textsc{GTS})
code \cite{WangGTS2006,WangGTS2010}. The \textsc{GTS} code performs
nonlinear gradient-driven electrostatic turbulence simulations based
on a generalized gyrokinetic simulation model using a $\delta f$
particle-in-cell approach. The presented \textsc{GTS} simulations
of NSTX discharges $\#128453$ and $\#135388$ take into account a
comprehensive influence of a number of relevant physical effects,
including fully kinetic electrons, realistic geometry constructed
using experimental data as well as plasma profiles which are read
from TRANSP \cite{Hawryluk1980}. The global simulations cover a wide
region of normalized minor radii, from $r_{\mathrm{tor}}=0.2$ to
$0.8$ ($r_{tor}$ represents the square root of the toroidal flux
normalized with its value at the separatrix). Convergence in marker
numbers was found for $80$ particles per cell per species. The spatial
grid size in the perpendicular direction is approximately equal to
or less than the local ion gyroradius $\rho_{\mathrm{i}}$. The wavenumber
range that is simulated is $k_{\perp}\rho_{\mathrm{i}}\lesssim2$,
which covers the typical low-$k$ turbulence due to ion temperature
gradient (ITG) mode and trapped electron mode (TEM).

\selectlanguage{american}%
\begin{widetext}

\begin{figure}[H]
\selectlanguage{english}%
\begin{centering}
\includegraphics[scale=0.5]{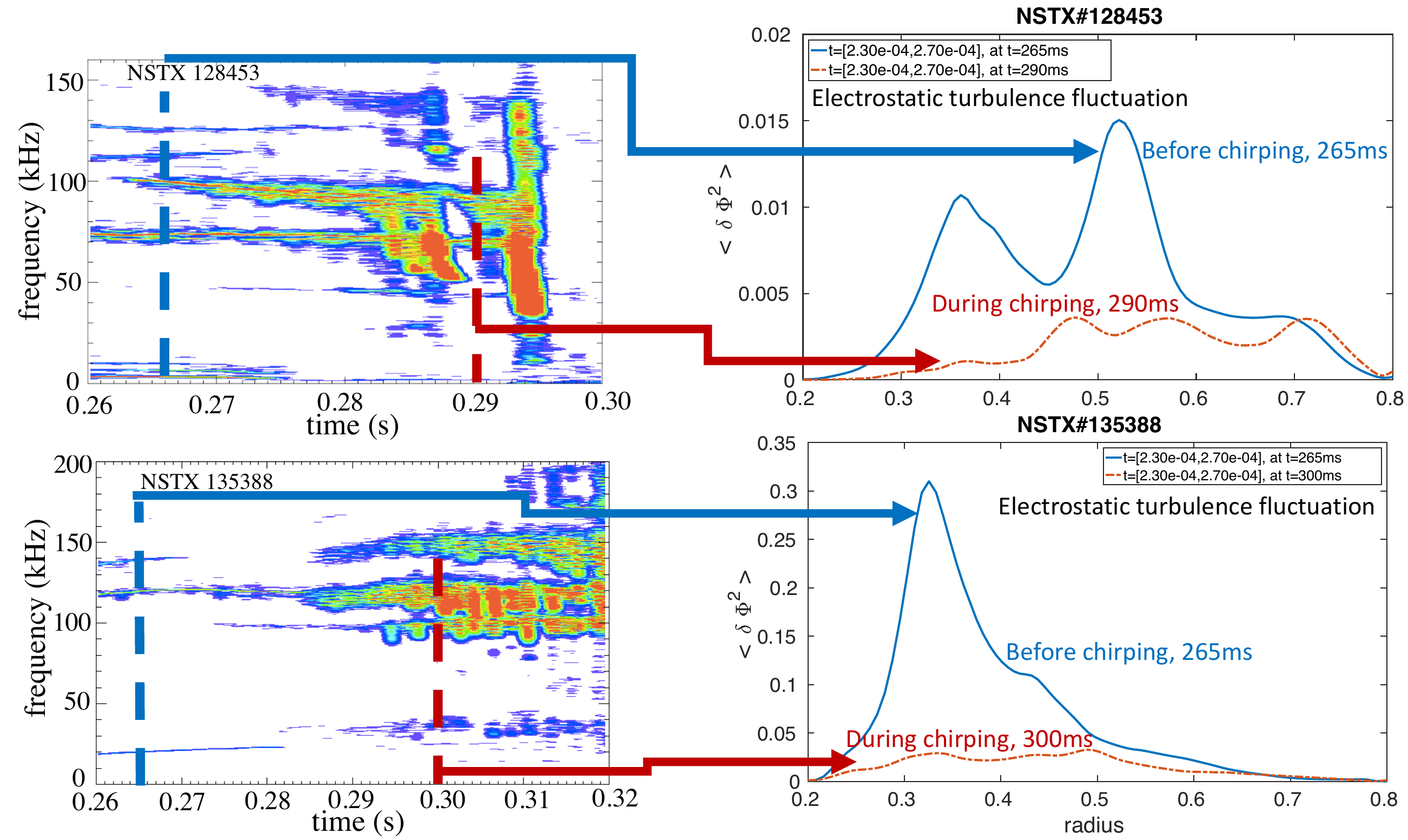}
\par\end{centering}
\selectlanguage{american}%
\caption{\foreignlanguage{english}{Correlation between low anomalous transport and Alfvénic chirping,
as calculated by GTS global gyrokinetic simulations, for NSTX pulses
$\#128453$ and $\#135388$.\label{fig:GTSCorr}}}
\end{figure}

\end{widetext}

\begin{figure}
\selectlanguage{english}%
\begin{centering}
\includegraphics[scale=0.3]{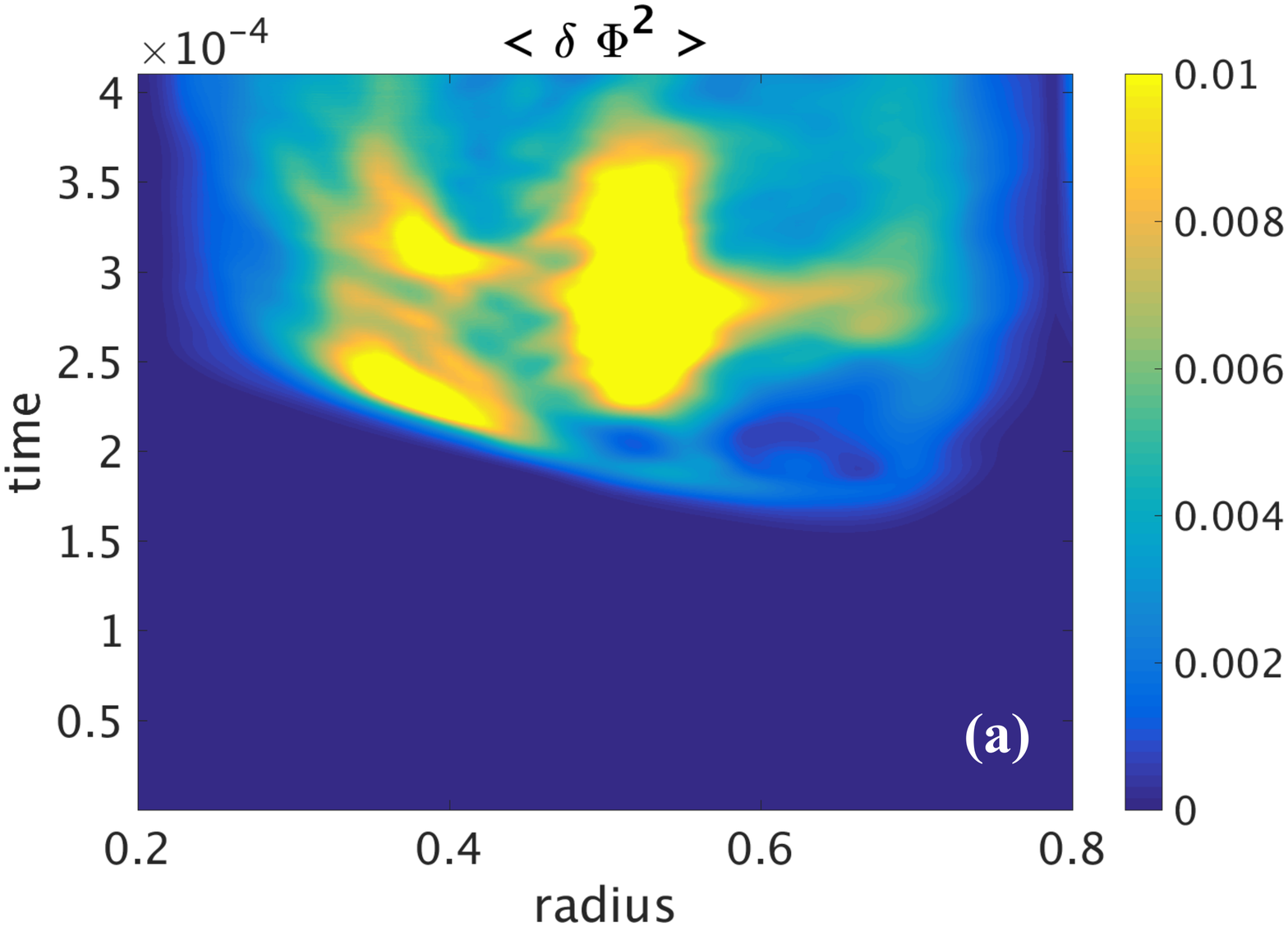}
\par\end{centering}
\begin{centering}
\includegraphics[scale=0.3]{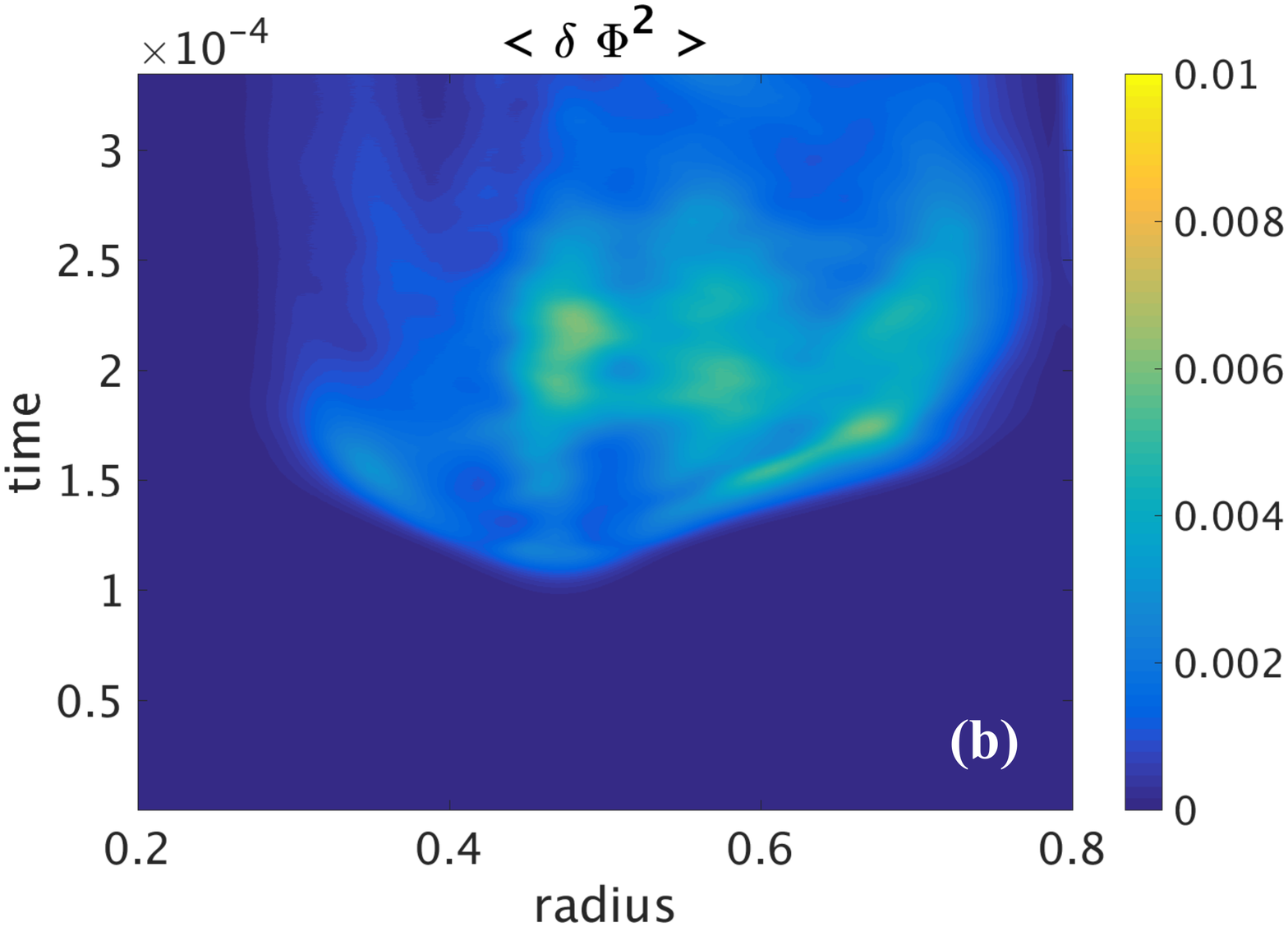}
\par\end{centering}
\selectlanguage{american}%
\caption{\foreignlanguage{english}{Averaged squared electrostatic potential fluctuation normalized with
the square of the ion temperature at mid-radius (indicated by the
color code), as a function of time, expressed in units of $10^{-4}s$
and the minor radius, expressed in terms of the square root of the
toroidal flux divided by the toroidal flux at the edge. The results
were obtained by the GTS code for the time slices (a) before chirping
starts, at $t=265ms$ and (b) during the chirping, at $t=290ms$,
for NSTX discharge $\#128453$. The color code scale is the same for
both plots, which indicates a substantial reduction of the turbulent
activity from (a) to (b).\label{fig:Electrostatic-turbulent-fluctuat-128453}}}
\end{figure}

\begin{figure}
\selectlanguage{english}%
\begin{centering}
\includegraphics[scale=0.3]{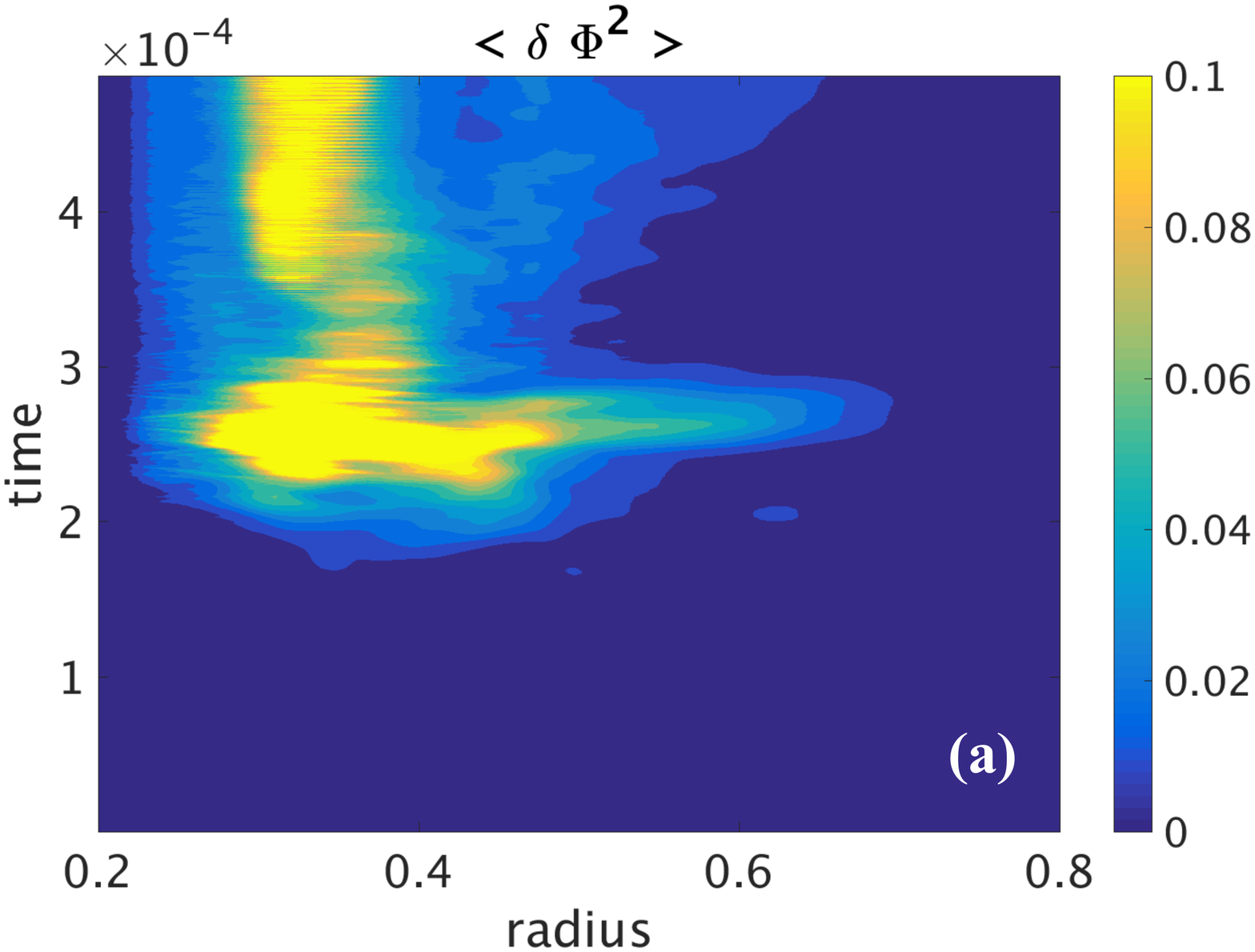}
\par\end{centering}
\begin{centering}
\includegraphics[scale=0.3]{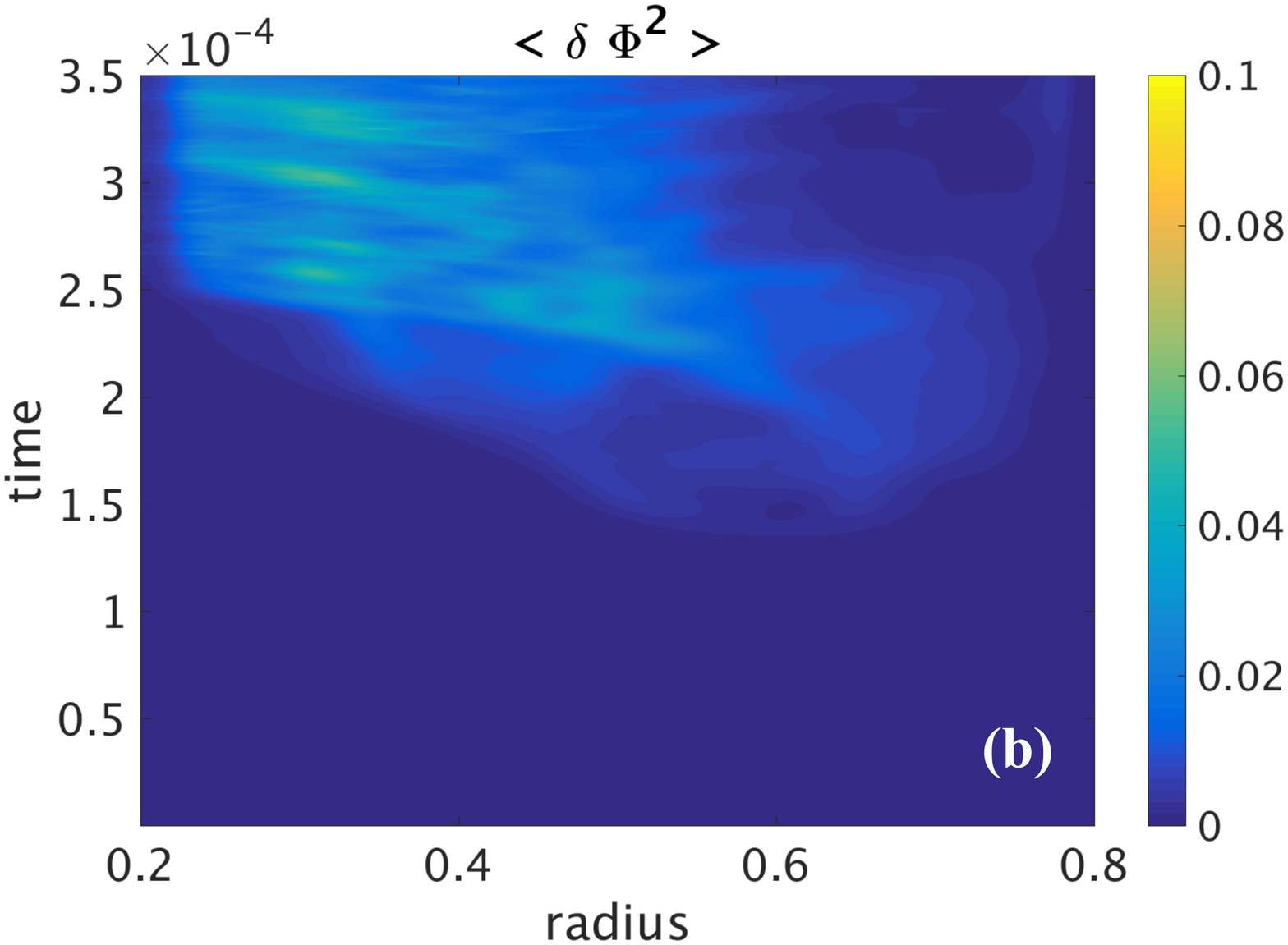}
\par\end{centering}
\selectlanguage{american}%
\caption{\foreignlanguage{english}{Averaged squared electrostatic potential fluctuation normalized with
the square of the ion temperature at mid-radius (indicated by the
color code), as a function of time, expressed in units of $10^{-4}s$
and the minor radius, expressed in terms of the square root of the
toroidal flux divided by the toroidal flux at the edge. The results
were obtained by the GTS code for the time slices (a) before chirping
starts, at $t=265ms$ and (b) during the chirping, at $t=300ms$,
for NSTX discharge $\#135388$. The color code scale is the same for
both plots, which indicates a substantial reduction of the turbulent
activity from (a) to (b).\label{fig:Electrostatic-turbulent-fluctuat-135388}}}
\end{figure}

\begin{figure}
\selectlanguage{english}%
\begin{centering}
\includegraphics[scale=0.3]{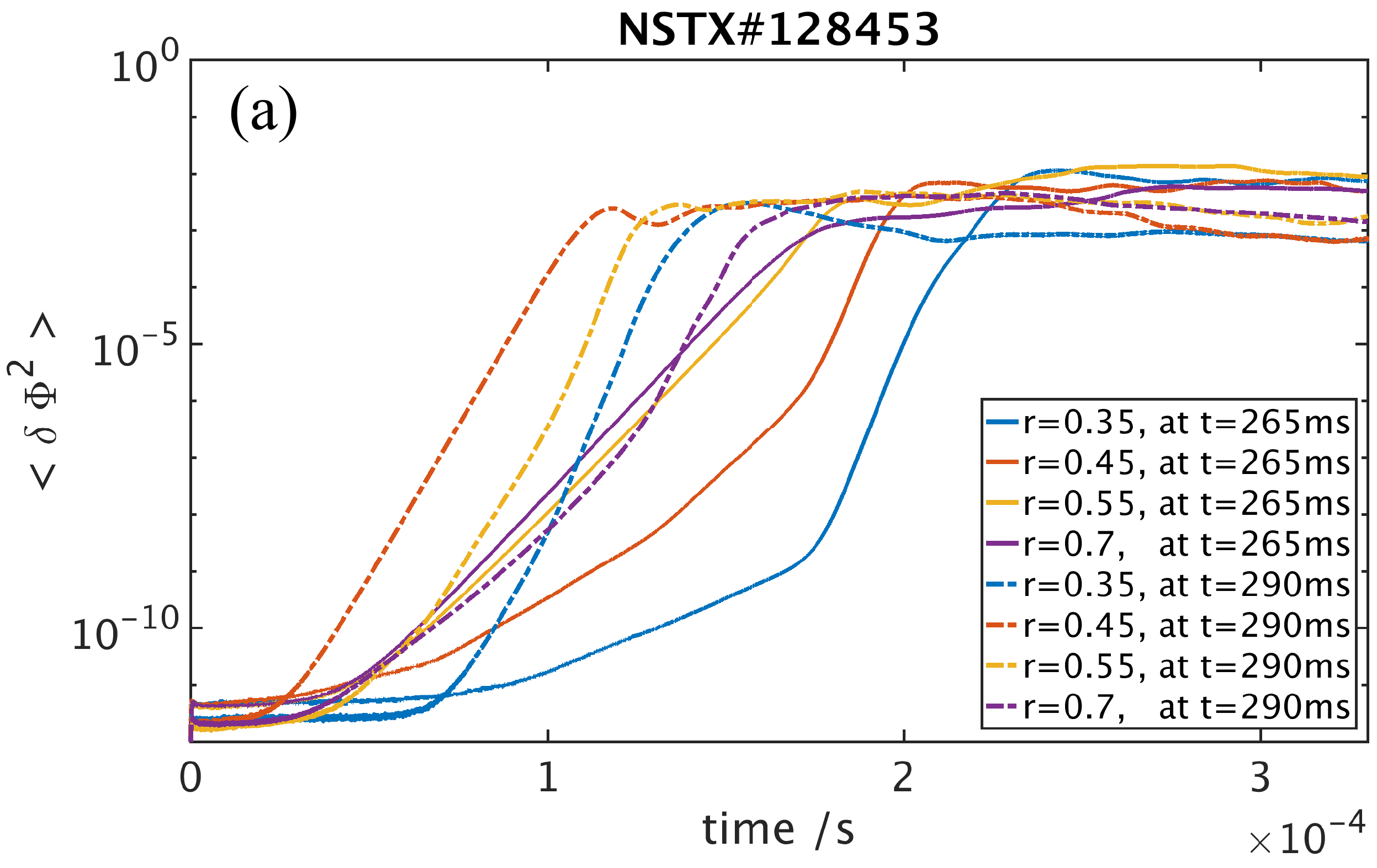}
\par\end{centering}
\begin{centering}
\includegraphics[scale=0.3]{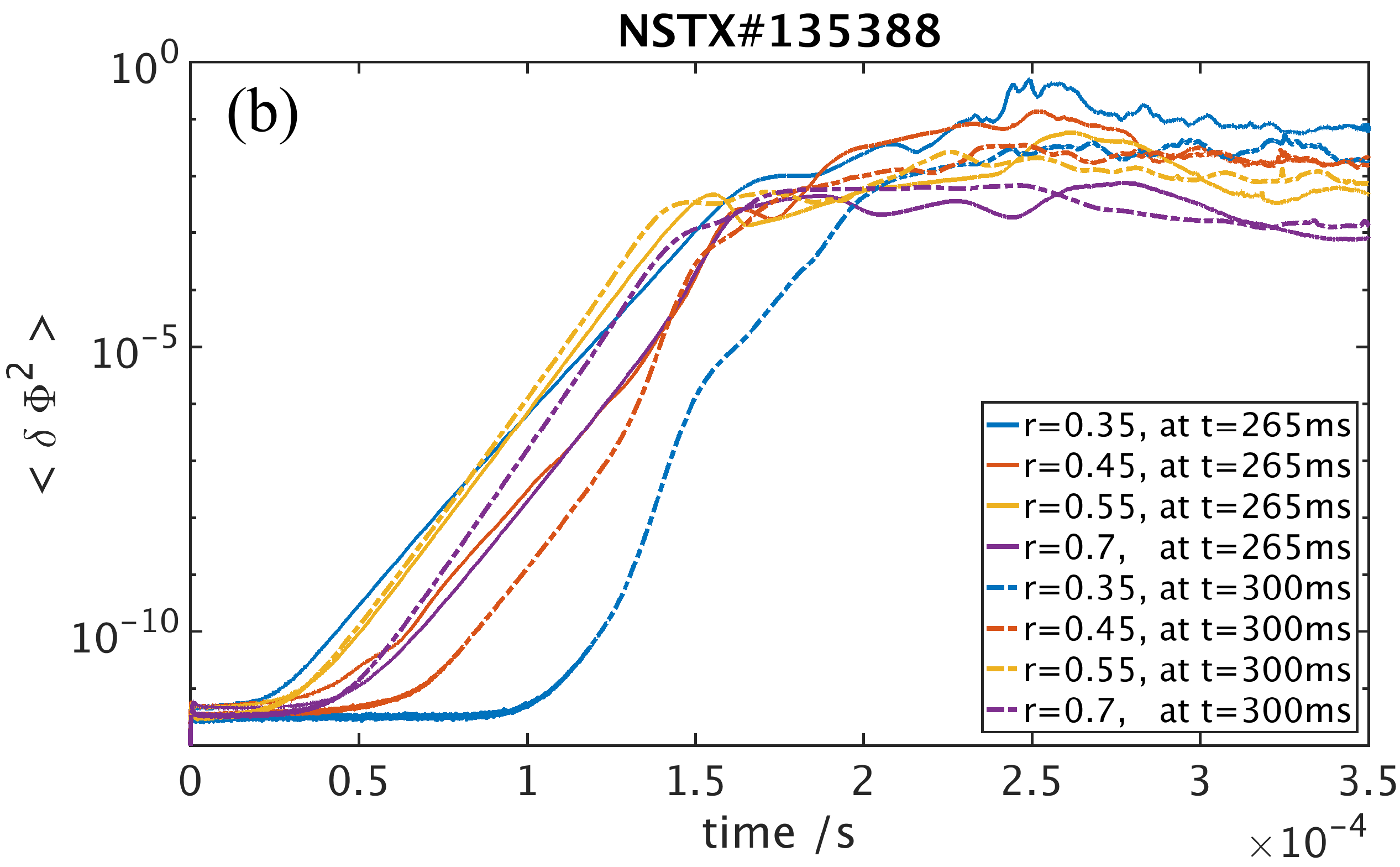}
\par\end{centering}
\selectlanguage{american}%
\caption{\foreignlanguage{english}{GTS computation of the time evolution of the averaged squared electrostatic
potential fluctuation normalized with the square of the ion temperature
at mid-radius, as a function of time for NSTX discharges (a) $\#128453$
and (b) $\#135388$, before (continuous curves) and during (dashed
curves) the chirping. The colors indicate several radial positions.\label{fig:LogPlotAmplitEvol}}}
\end{figure}

\selectlanguage{english}%
The spatiotemporal evolution of turbulence intensity, defined as $\left\langle \delta\Phi^{2}\right\rangle \equiv\langle(e\delta\phi/T_{\mathrm{i}})^{2}\rangle$,
with $\delta\phi$ being the electrostatic potential fluctuation,
$T_{\mathrm{i}}$ the ion temperature at the reference radius $r_{tor}=0.5$
and $e$ the elementary charge, is displayed in Figs. \ref{fig:GTSCorr},
\ref{fig:Electrostatic-turbulent-fluctuat-128453} and \ref{fig:Electrostatic-turbulent-fluctuat-135388}.
For $\#128453$, the fully developed ITG/TEM turbulence spreads mainly
around $r_{tor}=0.36$ and $r_{tor}=0.52$. For $\#135388$, ITG turbulence
evolves between $r_{tor}=0.25$ and $0.6$ with a peak around $r_{tor}=0.32$.
The location of the peaks is consistent with the radial position with
strongest temperature profile gradients. Interestingly, for both cases,
the level of turbulence is remarkably reduced as the mode undergoes
transition to chirping, in agreement with the theory prediction and
the observation on DIII-D \cite{DuarteAxivPRL}. The evolution and
saturation of the potential perturbation are illustrated in logarithmic
scale in Fig. \ref{fig:LogPlotAmplitEvol}, using the equilibrium
profiles before and during the chirping.

\begin{figure}
\includegraphics[scale=0.3]{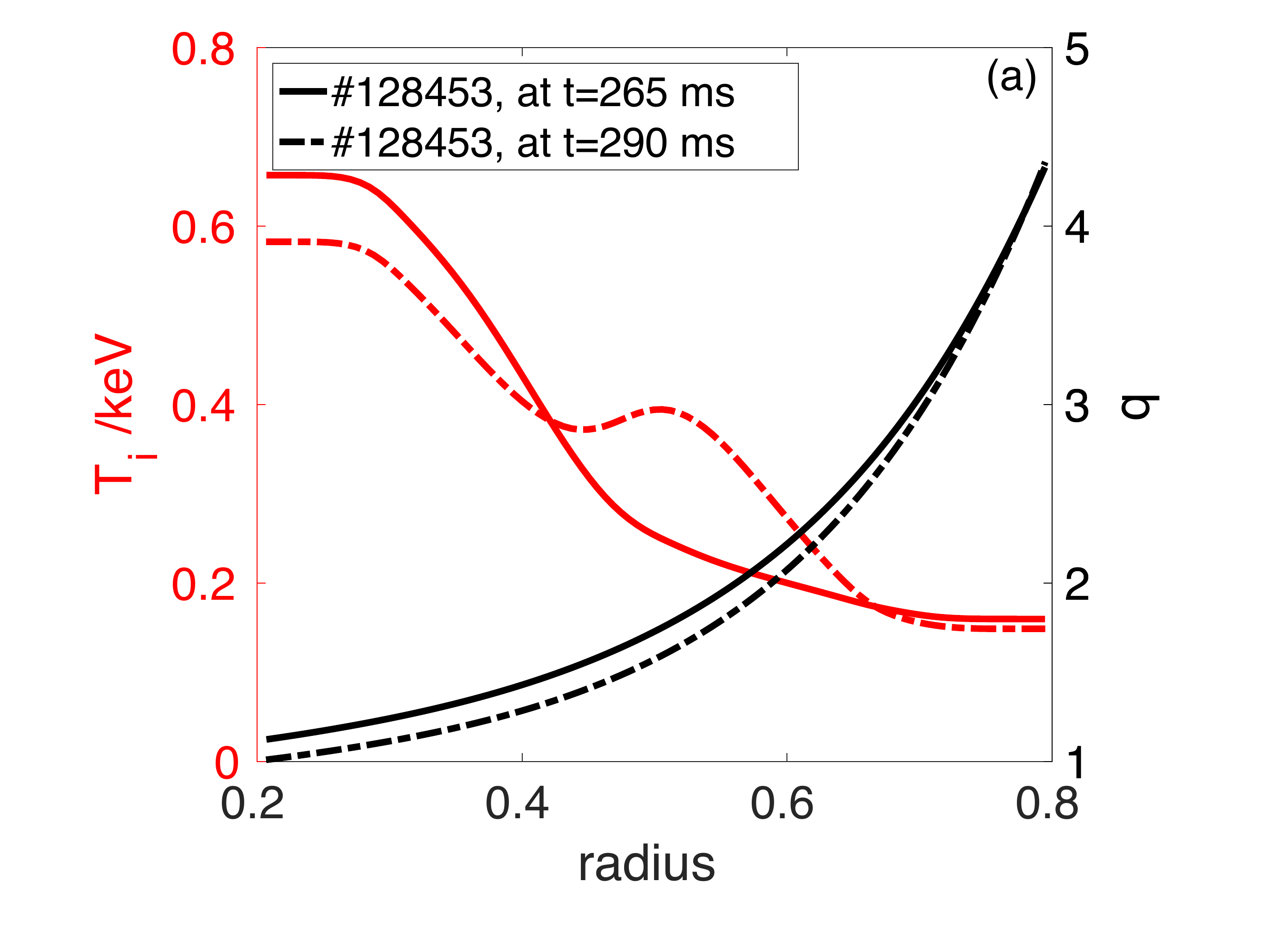}

\includegraphics[scale=0.3]{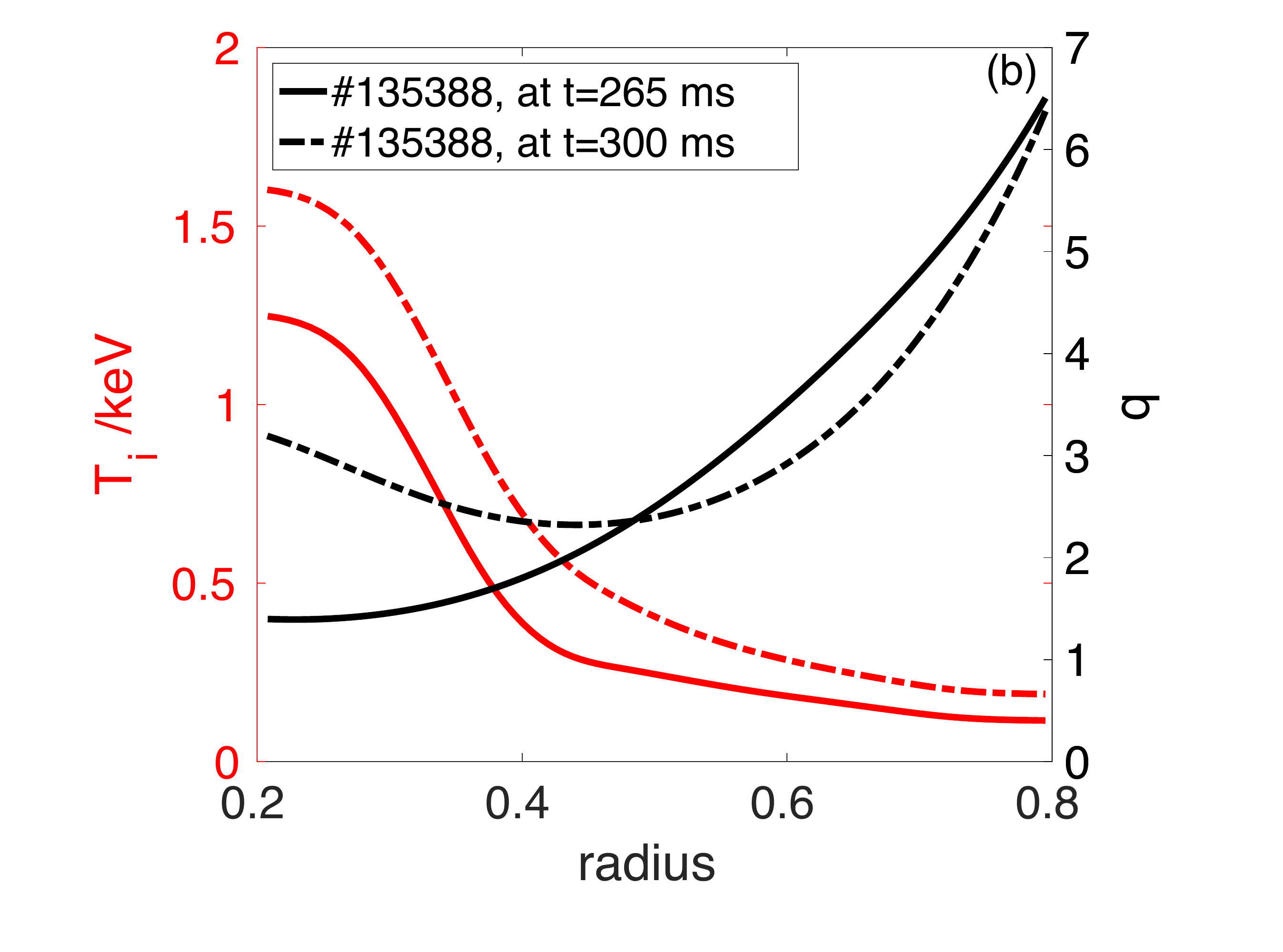}

\caption{\foreignlanguage{english}{Ion temperature profile (red) and safety factor profile (black), before
(continuous curves) and after (dot-dashed curves) the transition to
chirping. Simulations show that for shot 128453 (part (a)), the turbulence
drive is depleted via a flattening of $T_{i}$ at mid-radius while
for shot 135388 (part (b)), it is found that the reversal of $q$
has a stabilizing effect on the turbulence.\label{fig:QandTi}}}
\end{figure}

Interestingly, the turbulent levels in NSTX discharges $\#128453$
and $\#135388$ are found to be reduced due to different effects.
For $\#128453$, it is observed (Fig. \ref{fig:QandTi}(a))that the
ion temperature gets increasingly flattened in the core between $t=265ms$
and $t=300ms$, thus depleting the drive of the ITG modes. For $\#135388$,
the $q$ profile, which was monotonic at $t=265ms$, becomes reversed
at $t=290ms$ (Fig. \ref{fig:QandTi}(b)). The simulation result is
in agreement with the prediction that a reversed $q$ profile should
have an effect on the micro-turbulence supression \cite{Kishimoto1999}. 

The extraordinarily high electrostatic turbulence potential for case
$\#135388$ (as compared to $\#128453$) can be partially understood
when looking at one of the main driving sources, $\eta=\nabla T_{\mathrm{i}}/\nabla n_{\mathrm{i}}$.
In both time slices of the discharge NSTX $\#135388$, this value
is up to a factor of 5 to 6 higher than in the scenarios of NSTX $\#128453$.
Especially at lower radii (where the stabilizing shear is lower),
this high $\eta$ value is considered responsible for the high turbulence
levels. 

\selectlanguage{american}%

\begin{figure}
\includegraphics[scale=0.38]{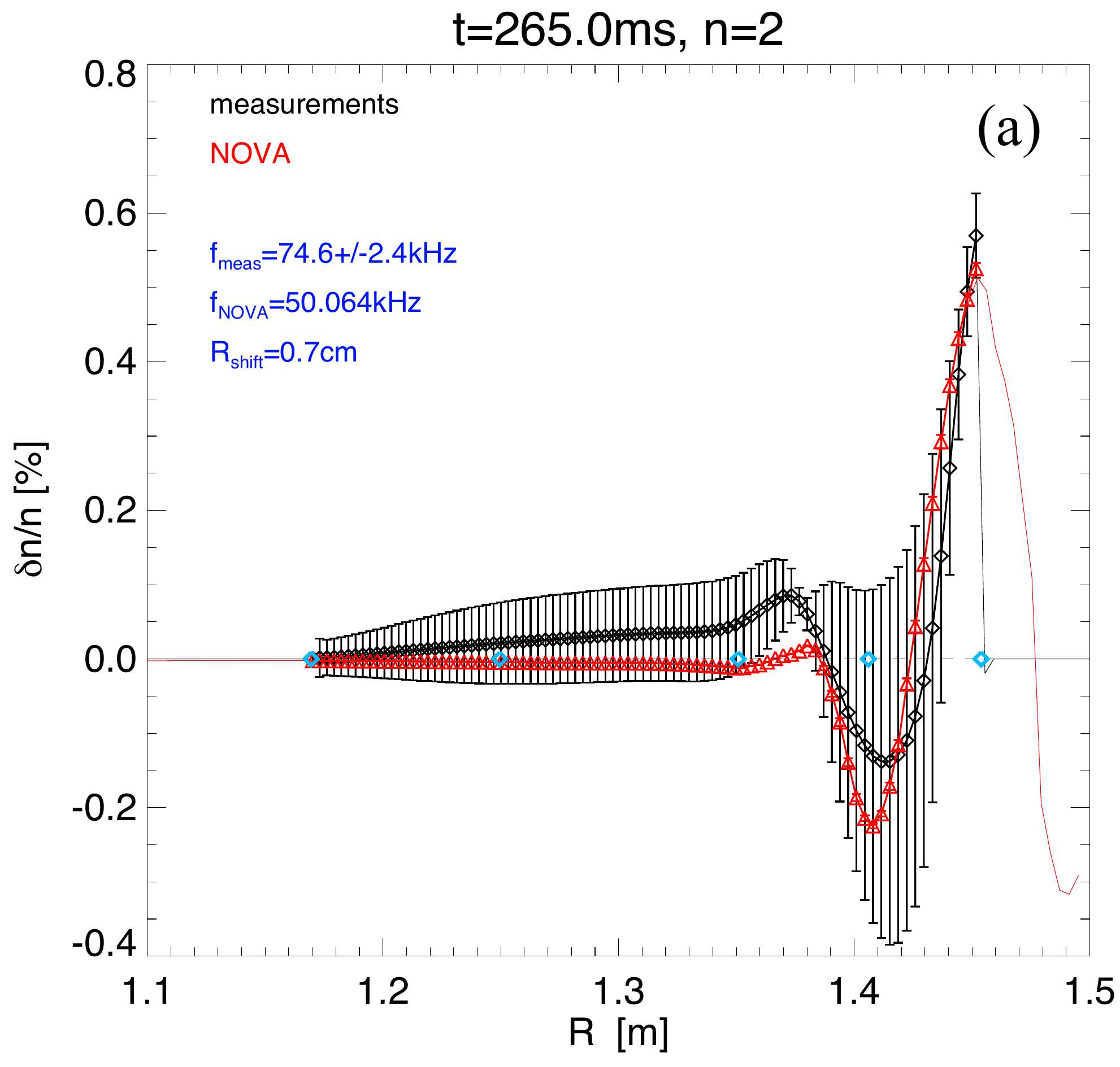}

\includegraphics[scale=0.38]{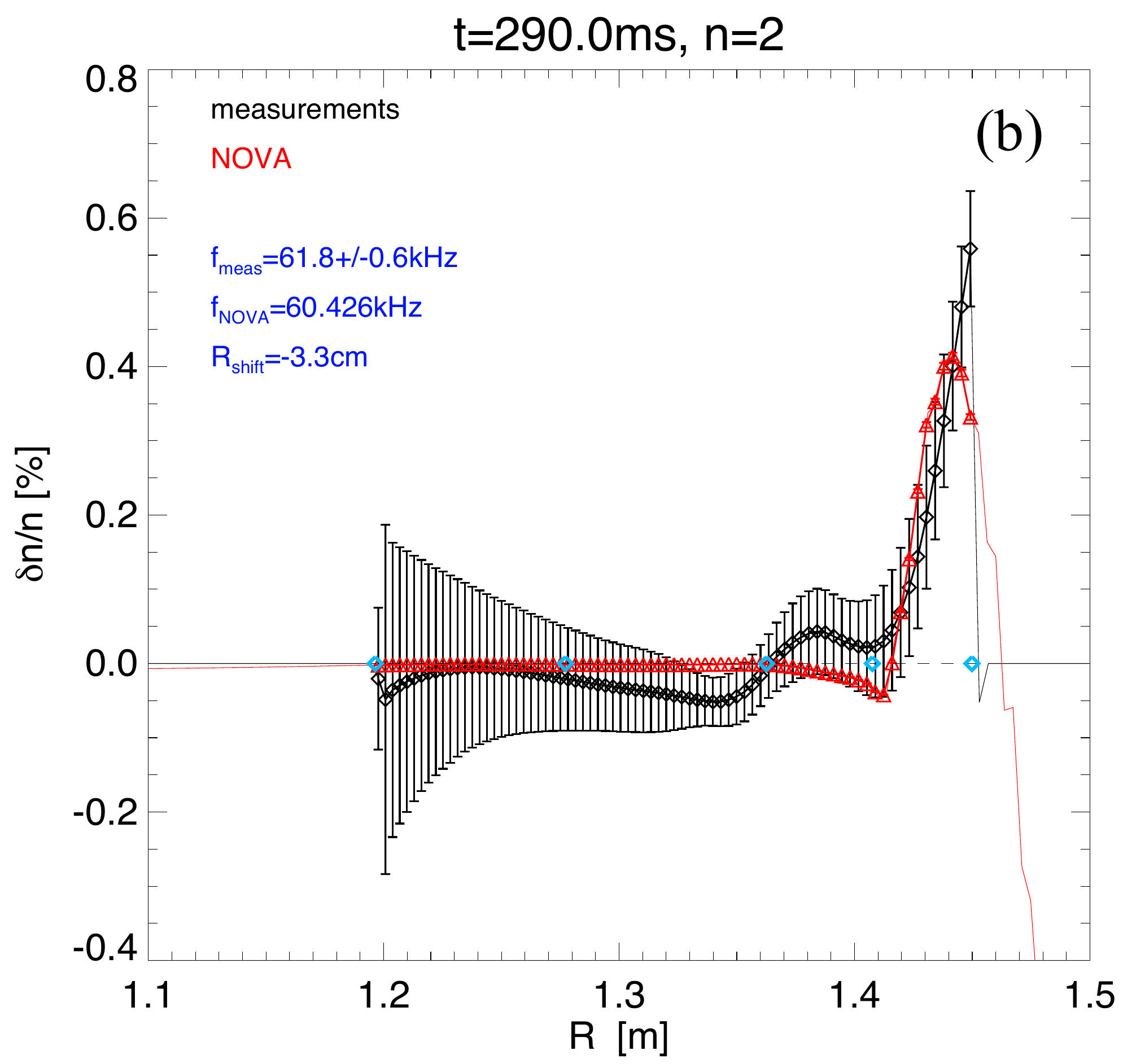}

\caption{\foreignlanguage{english}{Identification of mode structures for an $n=2$ mode NSTX shot 128453
calculated by NOVA (in red) compared with reflectometer measurements
(in black) for (a) $t=265ms$ and (b) $t=290ms$. $\delta n$ is the
perturbed density while $n$ is the background density. The reflectometer
measurement positions are shown by the cyan diamonds.\label{fig:ModeStruct}}}
\end{figure}

\selectlanguage{english}%
In order to evaluate the criterion for chirping likelihood proposed
in \cite{DuarteAxivPRL}, we first need to categorize the modes that
are measured. For this purpose, we employ the reflectometer data to
provide information regarding the mode structure, which can be compared
with the eigenmodes calculated by the NOVA code \cite{CHENG1985,Gorelenkov1999ChengFu}.
Due to lower density, for shot \#135388 only two channels of the reflectometer
could be used, which did not allow for proper mode identification.
The measured mode structure compared to its best match from NOVA results
is shown in Fig. \ref{fig:ModeStruct} for shot $\#128453$ before
and during the chirping phase.

In order to calculate the relative importance between fast ion stochasticity
arising from turbulent processes and collisional processes, we use
the approach introduced by Lang and Fu \cite{LangFu2011}, which in
the notation defined in \cite{DuartePoP2017}, is approximately given
by 
\begin{equation}
\frac{\nu_{turb}}{\nu_{scatt}}\backsimeq\left[\frac{D_{EP}\left(\frac{q_{EP}}{m_{EP}}\frac{\partial\psi}{\partial r}\right)^{2}}{2\nu_{\perp}R^{2}\mu B}\right]^{1/3}\label{eq:Ratio}
\end{equation}
where $\nu_{turb}$ and $\nu_{scatt}$ are the effective turbulent
and collisional scattering frequencies, $D_{EP}$, $q_{EP}$ and $m_{EP}$
are the energetic particle diffusivity, charge and mass, respectively.
$\psi$ is the poloidal magnetic flux divided by $2\pi$, $r$ is
the minor radius, $\nu_{\perp}$ is the $90^{o}$ pitch angle scattering
rate, $R$ is the major radius, $\mu$ is the magnetic moment and
$B$ is the magnetic field intensity. $D_{EP}$ is calculated using
the scalings of \cite{ZhangLinChen2008PRL}. We note, however, that
the \textsc{GTS} code is currently being extended to include a passive
energetic ion population. It should allow for more accurate estimates
of $D_{EP}$ in the future.

The weakly nonlinear mode dynamics can be described by an integro-differential,
time-delayed cubic equation for the mode amplitude \cite{BerkPRL1996,BerkPPR1997}.
The evolution to chirping has been identified to be associated with
the explosion of the solution of the cubic equation in a finite time
\cite{BerkCandy1999}. It has been shown that, for realistic tokamak
modes, a criterion for chirping likelihood $Crt$ \cite{DuarteAxivPRL}
should involve a phase-space integration over the multiple resonance
surfaces of a given mode. $Crt$ is sensitive to the ratio between
the effective frequencies due to stochastic ($\nu_{stoch}$) and coherent
($\nu_{coher}$) processes. We take $\nu_{stoch}=\nu_{scatt}+\nu_{turb}$
and $\nu_{coher}=\nu_{drag}$, where $\nu_{drag}$ is the effective
frequency associated with collisional drag (slowing down). For the
chirping criterion $Crt$ evaluation purpose, we use the kinetic postprocessor
NOVA-K \cite{CHENG1992,Gorelenkov1999Saturation}. We find that at
$265ms$, $Crt=+0.0027$ and at $290ms$, $Crt=-0.34$, which is consistent
with the experiment ($Crt<0$ implies more probability of chirping
while $Crt>0$ predicts that the wave will likely oscillate at a constant
frequency \cite{DuarteAxivPRL}). 

\selectlanguage{american}%

\section*{Predictive studies for ITER baseline scenarios}

ITER will employ two negative-ion-based neutral beam injection (NBI)
sources, which will account for $33MW$ of injected power \cite{HemsworthITER_NNBI_2009_NF,SinghITER_NBI_2017}.
Both the $3.5MeV$ fusion-born alpha particles and the tangentially
injected $1MeV$ NBI ions will have supra-Alfvénic velocities, allowing
them to interact with TAEs via their main resonance. An upper limit
of $5\%$ of fast ion loss has been established for ITER to sustain
a burning plasma \cite{JacquinotITERExpertGroup1999}. Therefore,
to understand the relevant mode evolution and fast ion transport due
to Alfvénic instabilities in ITER, it is instructive to anticipate
whether the modes will be more prone to have their frequencies locked
to the background equilibrium or be subject to rapid chirps \cite{SharapovNF2013}.
In recent years, a number of publications have addressed various aspects
of fast ion confinement in ITER, both linearly and nonlinearly \cite{LauberITERPPCF2015,PinchesITERPoP2015,Rodrigues_ITER_NF2015,SchnellerITER_PPCF2016,FitzgeraldITER_NF2016,Figueiredo_ITER_NF2016,RodriguesITER_NF2016,GorelenkovPPPLReport,GorelenkovNF2005ITER,VanZeelandNF2012_DIIID_ITER,VladNF2006_ITER,Isaev2017,Todo2014ITER}.
Most of these previous studies quantitatively addressed stability
and transport features. In this work we analyze another relevant aspect
of the problem, which is the prediction of the probable character
that Alfvénic waves will assume in ITER. This can be helpful in anticipating
the theoretical and numerical tools that may need to be developed
and employed. For example, for situations in which chirping is not
expected to take place, reduced quasilinear modeling \cite{Berk1995LBQ,GorelenkovDuarteNF2018,MengNF2018}
can be enough for a quantitative assessment of fast ion redistribution
due to Alfvénic modes.

ITER plasma performance can change considerably depending on the relative
external heating power between NBI, ion cyclotron resonant heating
(ICRH) and electron cyclotron resonant heating (ECRH). Heating mixes
with high NBI power increase the toroidal rotation and the fusion
yield but have the disadvantage of triggering Alfvénic instabilities
\cite{Budny_NF_2009_ITER_differentmixes}. Therefore it is necessary
to consider Alfvénic spectral behavior for the main three scenarios
\cite{Sips_ITER_scenarios} and make use of different plasma profiles
that the TRANSP code predicts for them. The first scenario is a reversed
shear, advanced (steady state) scenario in which most of the plasma
current is driven non-inductively (bootstrap current), with the \foreignlanguage{english}{current
density peak displaced from the center and a non-monotonic safety
factor $q$ profile. The second scenario is an elmy H mode, in which
$q$ is minimum (with a value just below $1$) at the center.} Finally,
the third scenario is a hybrid one, \foreignlanguage{english}{that
has weak or low magnetic shear and the current, although modified
externally, does not completely rely on non-inductive mechanisms.}

\begin{figure}
\selectlanguage{english}%
\includegraphics[scale=0.36]{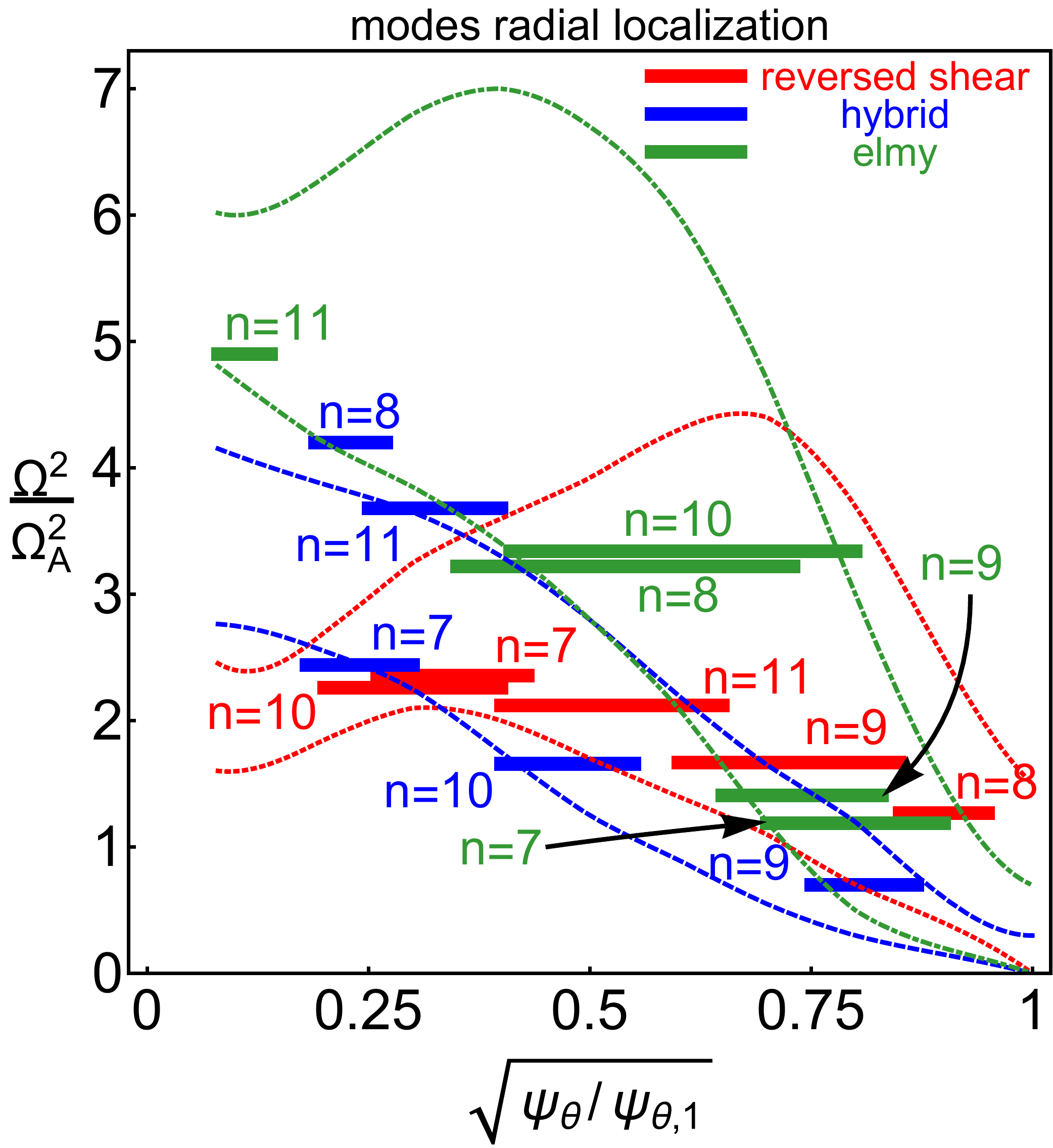}

\selectlanguage{american}%
\caption{\foreignlanguage{english}{Radial localization and frequency of the TAEs and RSAEs used in the
chirping prediction simulations of Fig. \ref{fig:ITERCritPlots}.
Dotted (red), dashed (blue) and dot-dashed (green) curves indicate
the envelope of the continuum for reversed-shear, hybrid and H mode
elmy cases. For the purpose of obtaining the envelope, a high toroidal
mode number $n=30$ was used. $\varPsi_{\theta}$ is the poloidal
magnetic flux and $\varPsi_{\theta,1}$ is its value at the plasma
separatrix. The modes angular frequency $\Omega$ is nomalized with
$\Omega_{A}=v_{A,axis}/(q_{1}R_{0})$, where $v_{A,axis}$ is the
Alfvén speed at the magnetic axis, $q_{1}$ is the safety factor at
the edge and $R_{0}$ is the major radius at the geometric center.
\label{fig:ITERModeStruct}}}
\end{figure}

\begin{figure}
\includegraphics[scale=0.35]{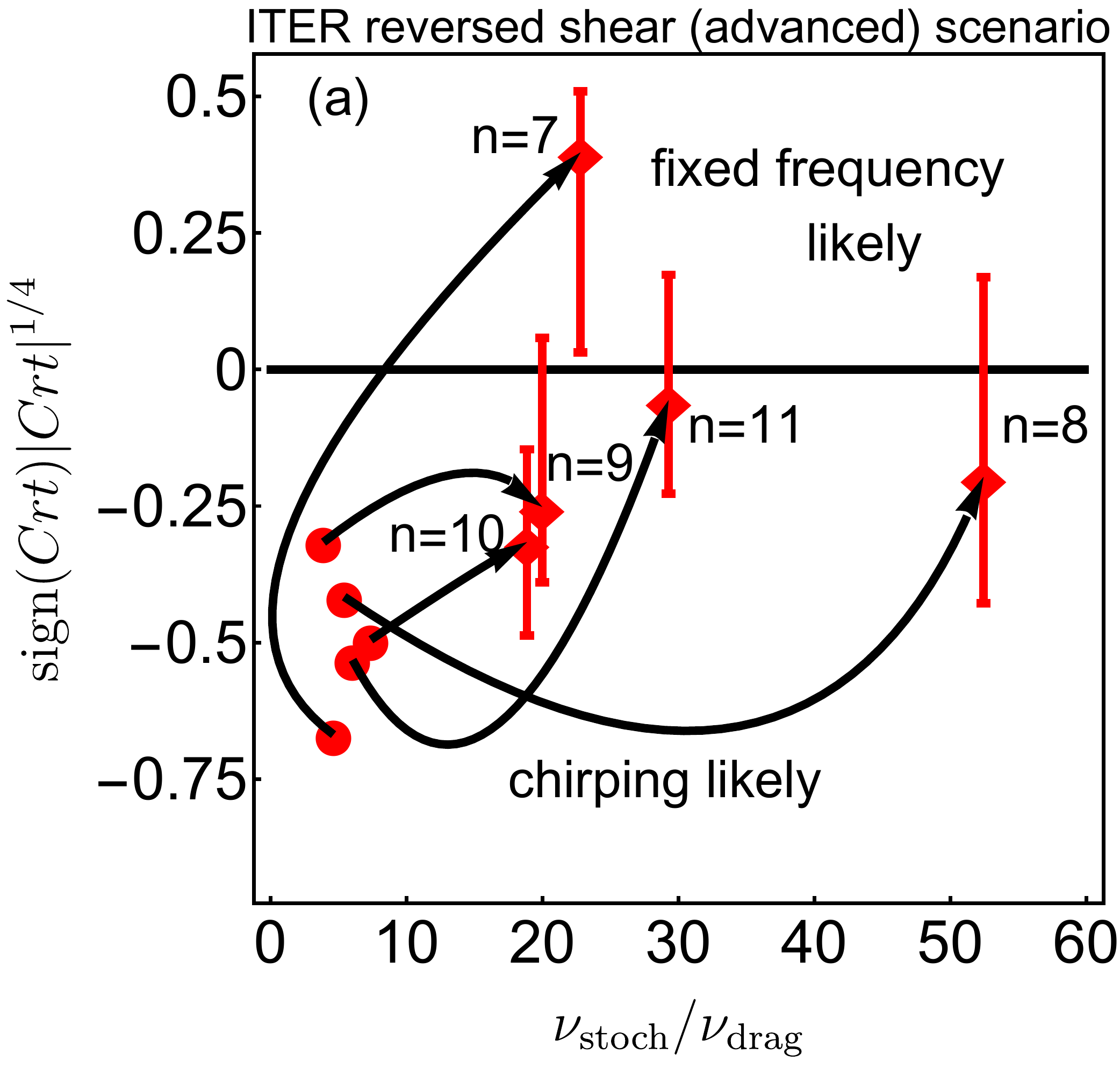}

\includegraphics[scale=0.33]{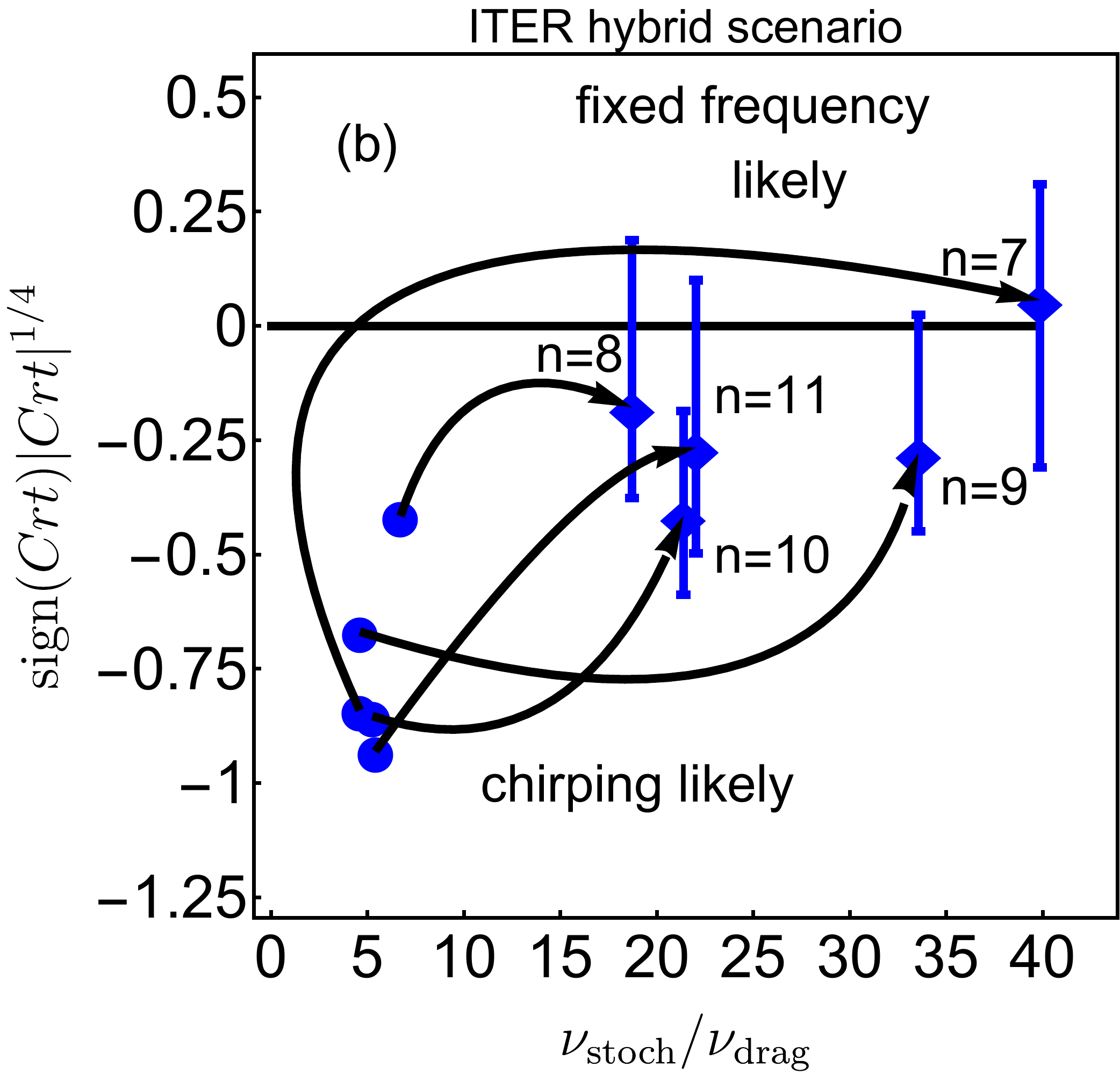}

\includegraphics[scale=0.33]{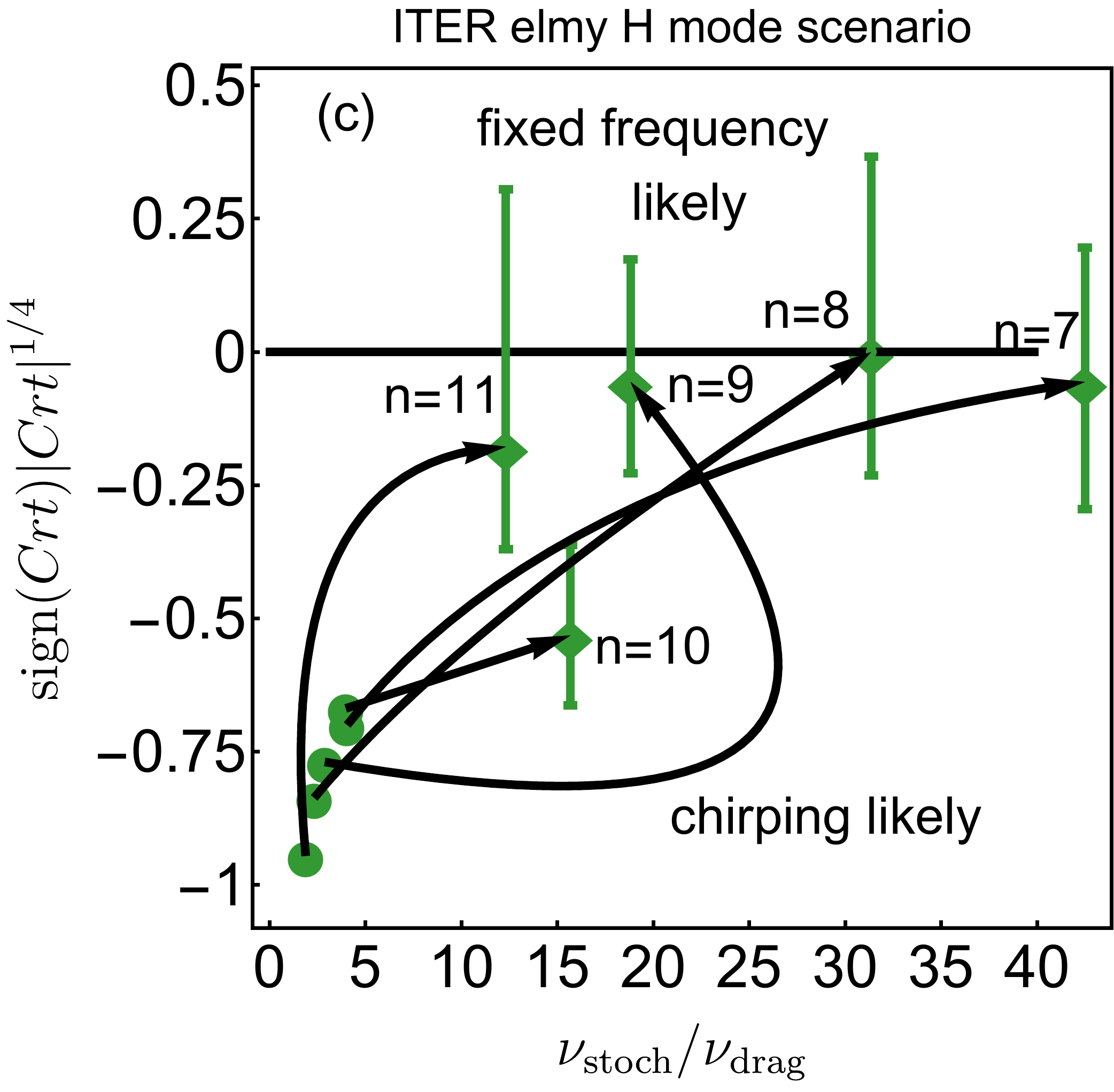}\caption{\foreignlanguage{english}{Evaluation of the chirping criterion $Crt$ \cite{DuarteAxivPRL,DuartePoP2017}
for the baseline scenarios (a) reversed shear, (b) hybrid and (c)
elmy h mode. It is shown the prediction for individual modes with
(arrowhead diamonds) and without (arrow tail disks) the inclusion
of micro-turbulence stochasticity in the model. The arrows do not
describe any meaningful path but simply connect two points corresponding
to the same mode.\label{fig:ITERCritPlots}}}
\end{figure}

We utilize plasma parameters from a previous TRANSP/TSC analysis \cite{GorelenkovPPPLReport},
requiring $Q\geq10$ (where $Q$ is the ratio between the power generated
by fusion reactions and the power input to the plasma). We employ
the mode structures and resonances previously reported in \cite{GorelenkovPPPLReport},
which had their linear stability already assessed. For each ITER scenario,
we take the most linearly unstable modes and limit ourselves to toroidal
mode numbers $n$ from $7$ to $11$.

\selectlanguage{english}%
In order to be able to account for distinctive turbulence levels for
each of the three baseline scenarios, we use the results of Ref. \cite{ZhangLinChen2008PRL}
to scale the fast ion diffusivity with the thermal ion one, which
is obtained with TRANSP. \foreignlanguage{american}{We note that Albergante
et al \cite{Albergante2009PoP,Albergante2010,AlbergantePPCF2011}
used the GENE code \cite{JenkoGENE2000} to study the diffusivity
of fast ions due to ITG turbulence specifically for ITER-relevant
scenarios \cite{JacquinotITERExpertGroup1999,Sips_ITER_scenarios}.
An alternative for our study would be to use the computed fast ion
diffusivity in terms of the parallel and perpendicular components
of the velocity, shown in Fig. 6 of Ref. \cite{Albergante2009PoP}.
We note that the figure was produced for the specific case of $E=30T_{e}$.
However, to make an estimate less approximate, one can use the result
in Fig. 5b of Ref. \cite{Albergante2009PoP} that shows the dependence
of the diffusivity on the temperature, for the situation when the
velocity-space is integrated over, where it can be inferred that $D_{EP}\propto T_{i}^{-1}$.
The latter expression can be used as a correction factor to the diffusivity
resolved in velocity for the case $E=30T_{e}$, in order to have a
better estimate for the specific fast ion energy and background temperature
for each specific ITER case. Interestingly, we find that the scalings
found in Ref. \cite{ZhangLinChen2008PRL} provide values for $D_{EP}$
reasonably compatible with the ones extracted from Albergante's work,
for most cases examined here.}

\selectlanguage{american}%
The frequency and radial localization of the 15 modes used in the
analysis are shown in Fig. \ref{fig:ITERModeStruct}, overlaid with
the envelopes of the continua. These envelopes were calculated by
interpolating the tips of each continuum curve, using high values
of $n$ in order to be able to produce more reliable curves, representative
of all modes of a given scenario.

The fact that the TAEs and the RSAEs in ITER can be driven by both
alphas and beam ions is taken into account in the analysis via a weighted
average of the individual contributions entering Eq. \ref{eq:Ratio}.\foreignlanguage{english}{
The predictions for the three baseline scenarios for ITER are shown
in Fig. \ref{fig:ITERCritPlots}. They indicate that most of the unstable
TAEs and RSAEs are located close to the boundary between the chirping
and steady frequency regions.} We note that without the addition of
micro-turbulent stochasticity in the model, all modes are predicted
to lie in the region that allows for chirping. However, upon the addition
of $\nu_{turb}$ to $\nu_{scatt}$ (represented by the arrows in Fig.
\ref{fig:ITERCritPlots}), most modes get very close to the borderline
$Crt=0$. Since there could be a considerable error in the estimate
of $D_{EP}$, we show error bars in Fig. \ref{fig:ITERCritPlots},
which indicate by how much the prediction for $Crt$ would change
if $D_{EP}$ is multiplied by two (upper bars) and divided by two
(lower bars). 

It is worth comparing the predictions for ITER with previous analyses
of DIII-D discharges \cite{DuarteAxivPRL}. Because of the inverse
dependence of $D_{EP}$ on the EP energy, we see that for ITER, $D_{EP}$
is about an order of magnitude smaller with respect to typical values
inferred for DIII-D. On the other hand, the Alfvén speed $v_{A}$
is about $2.5$ times larger in ITER because of higher field, which
means that the $90^{o}$ degree pitch angle scattering rate $\nu_{\perp}$
(see definition in Eq. 3 of Ref. \cite{DuartePoP2017}) at $v_{A}$
is also an order of magnitude smaller than in DIII-D. The combination
of the other parameters involved in the ratio \ref{eq:Ratio} do not
change appreciably in the comparison between the two tokamaks, therefore,
the ratio \ref{eq:Ratio} appears to be of the same order in DIII-D
and in ITER. 

Each ITER mode in Fig. \foreignlanguage{english}{\ref{fig:ITERCritPlots}}
appears to have its own threshold for steady/chirping transition.
Therefore it is challenging to draw general conclusions regarding
their spectral nature. It is however notorious that most modes appear
to be close to the borderline. This means that any additional stochastic
mechanism, such as RF heating, interaction with neoclassical tearing
modes (NTMs), field ripples, energy diffusion, mode overlap, would
contribute to make the modes cross towards the positive $Crt$ region.
ITER will be able to operate on a range of ICRH power, which can deliver
up to $20MW$ \cite{Budny_NF_2009_ITER_differentmixes}. It remains
to be understood how strong the effect of ICRH can be on the fast
ions that are in resonance with Alfvén waves.

\section*{Summary and conclusions}

We reported a study of rare transitions between mode constant frequency
and mode chirping in NSTX using global gyrokinetic simulation and
chirping criterion analysis. The results indicate that fast ion anomalous
diffusion likely mediates the transition, for cases when RF-induced
diffusion and changes in beam beta are not present. The chirping phase
is observed to be achieved following a marked decrease in the turbulent
field amplitude, which is consistent with the interpretation proposed
in Ref. \cite{DuarteAxivPRL}.

We have studied the likelihood of TAEs to exhibit chirping for typical
ITER scenarios (reversed shear, elmy and hybrid). A few modes were
found to be distinctly on either oscillation regime. However, most
modes were found to be borderline between the two phases. If additional
mechanisms for detuning fast ion from a resonance (e.g., induced by
3D fields, ICRH, NTMs, energy diffusion) will be important in ITER,
then most modes can be expected to oscillate at a constant frequency.
Each of these additional mechanisms, however, deserve a study on their
own, which is beyond the scope of this work. We also note that the
constraint $Q>10$ imposed by the analysis means that the thermal
diffusion could be too optimistic, which would increase the likelihood
for chirping. Besides, this constraint may mean that we are overestimating
the drive of the modes.

The analysis we presented in both parts of this work assumed that
the modes remain isolated throughtout their evolution. We note, however,
that because the toroidal mode numbers of Alfvénic waves in ITER will
be higher compared to present-day tokamaks, also higher poloidal mode
numbers will be important. This can lead to an enhanced phase-space
density of resonances (as an illustration, see Fig. 9 of \cite{SchnellerITER_PPCF2016}),
which can make resonance overlap more likely to occur. In the region
of overlap, resonant particles should move stochastically, which contributes
to destroy phase-space structures that sustain coherent phenomena,
such as chirping. In this case, our predictions with respect to the
nonlinear character of Alfvénic modes, which assumed isolated resonances,
are no longer valid. In the future, the description of the chirping
structures using more sophisticated numerical tools, such as guiding
center codes or initial value codes, could provide a more detailed
picture of the mode dynamics in realistic toroidal geometry. They
could be used as a comparison with the predictions from the present
reduced modeling effort.
\selectlanguage{english}%
\begin{acknowledgments}
This work was supported by the US Department of Energy (DOE) under
contract DE-AC02-09CH11466 and by the São Paulo Research Foundation
(FAPESP) under grants 2012/22830-2 and 2014/03289-4. The authors acknowledge
Z. X. Lu for pointing out the stabilizing effect of a reversed shear
profile, S. M. Kaye for preparing a TRANSP run for NSTX shot 128453
and S. Ethier for preparing the TRANSP profiles and equilibrium input
files for GTS. The GTS simulations for this work were run on the supercomputer
HYDRA, provided by the Max Planck Computing and Data Facility (MPCDF)
in Garching, Germany. Computational resources as well as support are
gratefully acknowledged. 
\end{acknowledgments}

\bibliographystyle{apsrev4-1}
\phantomsection\addcontentsline{toc}{section}{\refname}
%

\end{document}